\documentclass[
preprint,
 mathabx,amssymb,
 aps,
]{revtex4-2}
\usepackage{graphicx}
\usepackage{dcolumn}
\usepackage{bm}

\usepackage[utf8]{inputenc}
\usepackage{natbib}
\usepackage{float}
\usepackage{amsmath}
\usepackage[ruled,vlined,noresetcount]{algorithm2e}
\usepackage{caption}
\usepackage{subcaption}
\usepackage{fancyhdr}
\usepackage{lastpage}
\usepackage{subcaption,mathtools}
\usepackage{xcolor}
\usepackage{tikzducks}
\usepackage{upgreek,textgreek}
\expandafter\let\csname equation*\endcsname=\relax 
\expandafter\let\csname endequation*\endcsname=\relax

\DeclareSymbolFont{largesymbolsA}{U}{txexa}{m}{n}
\DeclareMathSymbol{\varprod}{\mathop}{largesymbolsA}{16}

\DeclareFontFamily{U}{mathx}{\hyphenchar\font45}
\DeclareFontShape{U}{mathx}{m}{n}{
      <5> <6> <7> <8> <9> <10>
      <10.95> <12> <14.4> <17.28> <20.74> <24.88>
      mathx10
      }{}
\DeclareSymbolFont{mathx}{U}{mathx}{m}{n}
\DeclareMathSymbol{\bigtimes}{1}{mathx}{"91}

\usepackage{bm}

\usepackage{graphicx}
\usepackage{soul}

\newtheorem{remark}{Remark}
\graphicspath{ {./figures/} }
\begin{document}

\title{Quantum estimation of the number of emitters for multiple fluorophores with the same spectral signature}

\author{Wenchao Li\textsuperscript{1}, Shuo Li\textsuperscript{2}, Timothy C. Brown \textsuperscript{3}, Qiang Sun\textsuperscript{2}, Xuezhi Wang\textsuperscript{1}, Vladislav V. Yakovlev\textsuperscript{4}, Allison Kealy\textsuperscript{5}, 
Bill Moran\textsuperscript{1,6}, Andrew D. Greentree\textsuperscript{2,*}}

\address{
1 School of Science, RMIT University, Melbourne, VIC 3001, Australia\\
2 ARC Centre of Excellence for Nanoscale BioPhotonics, RMIT University, Melbourne, VIC 3001, Australia\\
3 School of Mathematics, Monash University, Melbourne, VIC 3800, Australia\\
4 Department of Biomedical Engineering, Texas A\&M University, College Station, TX 77843, USA\\
5 School of Engineering, RMIT University, Melbourne, VIC 3001, Australia\\
6 Department of Electrical and
Electronic Engineering, University of Melbourne, VIC 3010, Australia}
\email{andrew.greentree@rmit.edu.au} 


\begin{abstract} 
Fluorescence microscopy is of vital importance for understanding biological function.  However most fluorescence experiments are only qualitative inasmuch as the absolute number of fluorescent particles can often not be determined.  Additionally, conventional approaches to measuring fluorescence intensity cannot distinguish between two or more fluorophores that are excited and emit in the same spectral window, as only the total intensity in a spectral window can be obtained.  Here we show that, by using photon number resolving experiments, we are able to determine the number of emitters and their probability of emission for a number of different species, all with the same measured spectral signature.  We illustrate our ideas by showing the determination of the number of emitters per species and the probability of photon collection from that species, for one, two and three otherwise unresolvable fluorophores.  The convolution Binomial model is presented to model the counted photons emitted by multiple species. And then the Expectation-Maximization (EM) algorithm is used to match the measured photon counts to the expected convolution Binomial distribution function. In applying the EM algorithm, to leverage the problem of being trapped in a sub-optimal solution, the moment method is introduced in finding the initial guess of the EM algorithm. Additionally, the associated Cram\'er-Rao lower bound is derived and compared with the simulation results. 
\end{abstract}
\maketitle

\section{Introduction}
Understanding complex biological function often requires precise localization of biological molecules in space and time to better understand their interactions and relevant chemical reactions and transformations. Cryogenic electron microscopy is capable of defining positions of atoms in complex molecules with angstrom accuracy \cite{Yip2020nature} but is often limited in capturing complex interactions of molecules in dynamic biological environment. Fluorescence and Raman microspectroscopies can provide structural and functional information about biological molecules; however, special measures are needed to extend the spatial resolution of optical imaging beyond the traditional diffraction limited spatial resolution defined by the excitation wavelength of light. The new generation of optical imaging methods based on super-resolution optical imaging, for which Nobel Prize in Chemistry was awarded in 2014, are gradually being adopted by the research community, and these techniques are now indispensable for fundamental understanding of biological function at the molecular level. However, many biological molecules are performing their function not along but in a coherent ensemble with other molecules. A typical example of such synergistic interaction is electron-transfer complex in mitochondrial membrane, where several cytochromes are involved in the ultimate production of ATP molecules which serve as a major energy fuel for a cell. To understand the function of mitochondrion and to assess its ability to efficiently produce ATP molecules, one needs to quantify the number of such cytochromes in membrane which is close to impossible using conventional methods. There are several challenges which are related to the size of the focal spot and internal dynamics of mitochondria which make it difficult to localize those different cytochromes within the membrane. Traditional approaches based on classical photon statistics have significant limitations due to either invasive nature, such as induced photobleaching which affects the electronic structure of biological fluorophores and amends its function, or complexity of signal collection and analysis \cite{Grubmayer2019methods}. On the other hand, quantum spectroscopy based on quantum statistics of detecting photons using photon resolved detectors and cameras \cite{Becerra2015np, Provaznik2020oe} (see also qCMOS by Hamamatsu \cite{hamamatsu2021whitepaper}) as we have shown recently \cite{Li2022oe}, makes it feasible to count individual emitters in a focal volume.

Techniques of fluorescence microscopy have become amongst the most used techniques for understanding biological function~\cite{Ellinger1929, Lichtman2005, Renz2013, Sanderson2014, Kubitscheck2017, Stockert2017}.  This typically involves   measurement of the uptake of functionalised fluorophores, or observation of the expression of fluorescent proteins in response to some stimulus~\cite{Lichtman2005, Renz2013, Sanderson2014, Kubitscheck2017, Stockert2017}. Fluorescence experiments are usually \emph{qualitative}, or at least relative, as the total number of fluorophores is often not knowable because  a conventional intensity measurement is unable to distinguish few bright emitters from many dim emitters~\cite{Lichtman2005}.  Moreover, if the fluorophores are excited and emit in the same spectral windows, then they may be impossible to distinguish with intensity only measurements. In \cite{Chon2021oe}, the average number of emitters in each species and the brightness ratio between multiple species are investigated and evaluated using high-order image correlation spectroscopy.

Earlier, we showed that the problem of quantitative determination of the number of emitters and the probability of photon detection could be solved for a single species of emitters with assumed identical properties~\cite{Li2022oe}.  Our approach there used the binomial distribution of the number of photons emitted in a pulsed fluorescence experiment.  By considering photon number resolving detectors (PNRD) ~\cite{Provaznik2020oe,Thekkadath2020thesis,Kardynal2008nphoton,Ma2017optica,Cahall2017optica,Schmidt2018LowTemPhys,Kalashnikov2011oe,Morais2020arXiv,Mattioli2016oe} we showed that the distribution of photons arriving in each detected pulse is uniquely specified by the number of emitters and the photon detection probability, so that these parameters can be determined with some confidence  given a particular measurement record.

Here we show that PNRD detection techniques can, in principle,  be used to discriminate between an arbitrary number of fluorescent species.  We assume that each species is defined by two parameters, the number of emitters, $M$, and the probability of detection $p$, which is assumed to be constant for each member of the species,  for instance,  by each species having a different transition dipole moment.  We assume no further ability to distinguish the species.

This paper is organised as follows.  We first introduce the measurement model and maximum likelihood estimation (MLE) approach.  We then discuss how the expectation-maximization (EM) technique is applied.  Finally we present simulations and Cram\'er-Rao lower bounds (CRLB) for the cases of one, two and three species.

\section{Measurement model and MLE}

\begin{figure}[htb!]
    \centering
    \includegraphics[width=.8\textwidth]{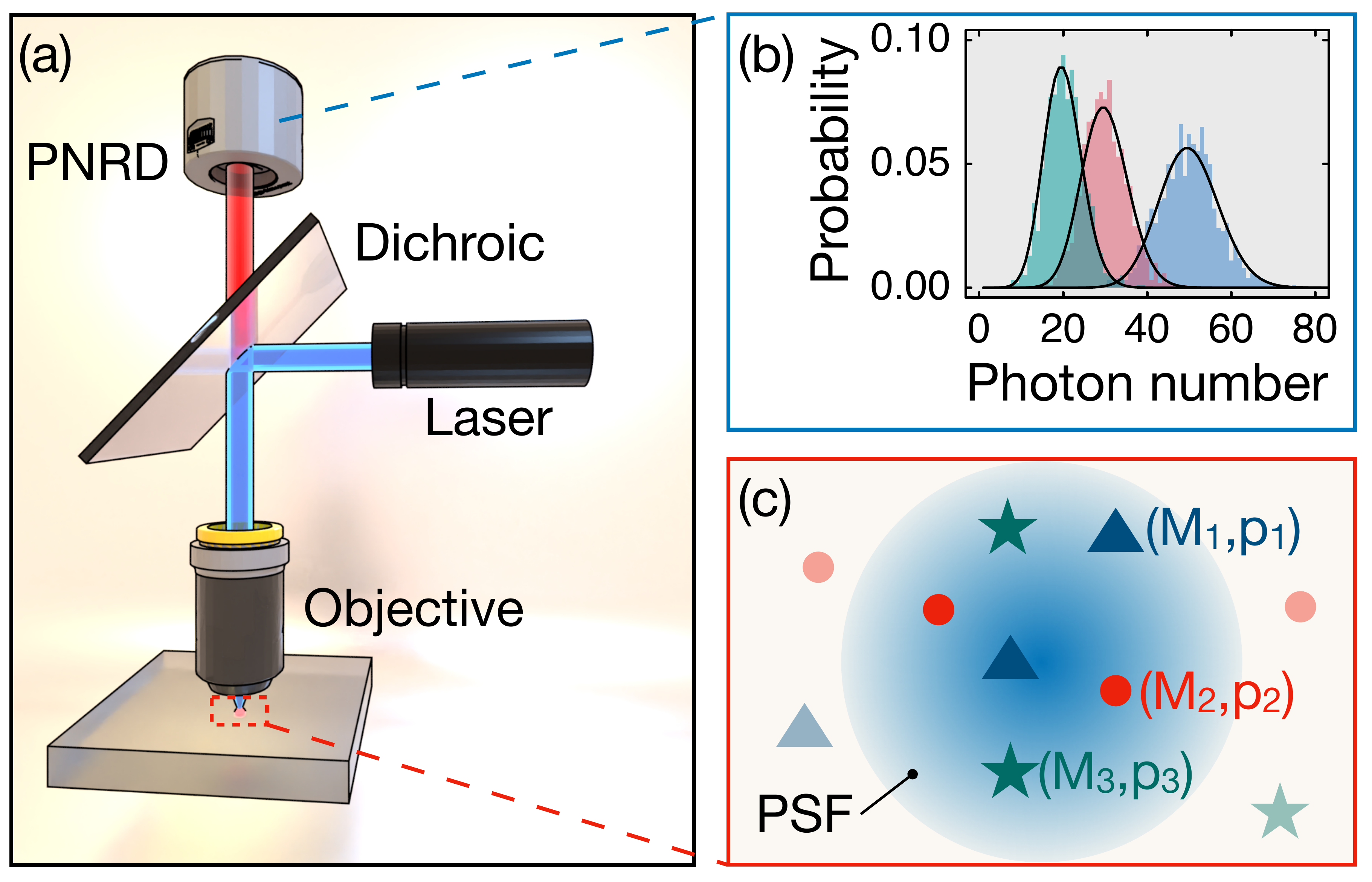}
    \caption{(a) Schematic of using PNRD to discriminate three species of emitters. The excitation laser (blue beam) passes through the dichroic mirror and focuses on three species $(M_1,p_1)$, $(M_2,p_2)$, $(M_3,p_3)$ under the diffraction limit locating within one Gaussian focal spot in (c). The emitting light (red beam) is detected by PNRD and the synthetic photon number resolving signal is shown in (b) as bar charts in three colors. With the measurement time increases the probability of detecting the photon numbers would follow Poisson distribution, shown as the black smooth curves. The area underneath each poisson curve or bar chart is always one, indicating that in a practical measurement the photon counts or the intensity from each species is the same, therefore they cannot be identified by conventional intensity-only based microscopy. However they generate distinguishing PNRD signals, building on which we introduce the MLE approach to tell them apart.}
    \label{fig:schematic}
\end{figure}

We consider an fluorescence experiment where there are $m$ distinct fluorophores (species);  species $j$ has $M_j$ emitters  with probability of photon detection from each member of  that species $p_j$.  A single excitation source (e.g. laser through microscope objective) excites the sample, and the fluorescence photons are collected by a PNRD. Schematic with four species emitters is shown in Fig.\ref{fig:schematic}.

This experiment is repeated a large number of times to build up statistics about the system. 
The number of fluorescence photons detected from each pulse, $y$ are counted at the outputs of the emitters; this is modeled as the sum of $m$ random variables sampled from  Binomial distributions with parameters $[M_j,p_j]$, $j=1,\cdots,m$, i.e.
\begin{align}
    Y=\sum_{j=1}^m Y_j,\;\;\;\text{where  }Y_j\sim {B}(M_j,p_j)\label{rv}
\end{align}
The probability mass function (PMF) of $Y=y$, $y=0,\cdots,M$ and $M=\sum_{i=1}^mM_i$, given parameters $\bm{\theta}=\left[M_1,p_1,M_2,p_2,\cdots,M_m,p_m\right]$ can be seen, by the law of total probability, to be a convolution of $m$ Binomial distributions, $pr_j(y_j|M_j,p_j)$ for $j=1,\cdots,m$ and $y_j=0,1,\cdots,M_j$, and so can be derived as
\begin{align}
    pr(Y=y|\bm{\theta})=\sum_{ \mathbf{y}\in\mathcal{Y}_y}\left(\prod_{j=1}^m\frac{M_j}{(M_j-y_{j})!y_{j}!}p_j^{M_j}(1-p_j)^{M_j-y_{j}}\right)\stackrel{\triangle}{=}\sum_{ \mathbf{y}\in\mathcal{Y}_y}\left(\prod_{j=1}^mpr_j(y_j|M_j,p_j)\right),\label{pdf}
\end{align}
where $\mathbf{y}=\left[y_{1},\cdots,y_{m}\right]$,  $\mathcal{Y}_y=\left\{\left[y_1,\cdots,y_m\right]\Big|\sum_{j=1}^my_j=y,y_j\in\left[0,1,2,\cdots,M_j\right]\right\}$. $\mathcal{Y}_y$ is actually the collection of partition of the integer $y$ into exactly $m$ parts.  

The fundamental problem of interest is to estimate $\bm{\theta}$ based on the measurements.  We show that this problem can be formulated as an MLE. In~\cite{shah1973distribution}, it is  shown that the probability of the sum of $m$ independent integer-valued random variables (not necessarily identically distributed), i.e. $pr(Y=y|\bm{\theta})$, may be calculated using a recurrence relation. Furthermore, the PMF of \eqref{rv} can be approximated by a Gaussian distribution, $\mathcal{N}\left(\sum_{j=1}^mM_jp_j,\sum_{j=1}^mM_jp_j(1-p_j)\right)$. In~\cite{eisinga2013saddlepoint,butler2017distribution,ha2017krawtchouk}, other more accurate approximation methods, such as  saddlepoint approximation, Kolmogorov approximation, or  Krawtchouk polynomial approximation, are provided. Since all the mentioned methods are either in the form of a recurrence formula or otherwise have no closed form, they cannot be directly used in deriving the MLE for $\bm{\theta}$. Accordingly, we use Eq. \eqref{pdf} for investigating the MLE.

In each experiment we record the peak corresponding photon number from the PNRD signal as $i$, $i = 0,...N$ and $N<=M$. We count the occurrence of $i$ in a series of experiments as $C_i$, then the data from a series of experiments can be given by the  frequency distribution $\left[C_0,C_1,\cdots,C_{N}\right]$, and $\nu=\sum_{i=0}^{N}C_i$ is the total number of experiments. The log-likelihood function can then be expressed as 
\begin{align}
\ell(C_0\cdots,C_{N}|\bm{\theta})=\sum_{i=0}^{N}C_i\log\sum_{ \mathbf{y}\in\mathcal{Y}_i}\left(\prod_{j=1}^mpr_j(y_j|M_j,p_j)\right)
\stackrel{\triangle}{=}\sum_{i=0}^{N}C_i\log L(\mathbf{y}|\bm{\theta})\label{likelihood}
\end{align}
where, similar to $\mathcal{Y}_y$, $\mathcal{Y}_i$ is the collection of  partition of the integer $i$ into $m$ parts. Furthermore, we assume that $f(\mathbf{y}|\bm{\theta})=\prod_{j=1}^mpr_j(y_j|M_j,p_j)$.

The MLE of $\bm{\theta}$, $\hat{\bm{\theta}}=[\hat{M}_1,\hat{p}_1,\cdots,\hat{M}_m,\hat{p}_m]$, is
\begin{align}
\hat{\bm{\theta}}=\arg\max_{\bm{\theta}\in\varprod_{j=1}^m \left(\mathbb{Z}^+\times[0,1]\right)}\ell(C_0\cdots,C_{N}|\bm{\theta})\label{MLE}
\end{align}
where $\varprod$ is the Cartesian product. From \eqref{pdf} to \eqref{MLE}, the underlying estimation problem is formulated as a parameterised MLE problem; that is, seeking the set of parameters $\bm{\theta}$ in the parameter space which yield maximum likelihood $\ell(C_0\cdots,C_{N}|\bm{\theta})$ based on the observations. 

The solution to  \eqref{MLE} when $m=1$ was provided in~\cite{Li2022oe}, where the MLE is proved to be an effective estimator and the associated CRLB is derived.  However, when $m>1$, solving \eqref{MLE} directly is very inefficient and even computationally impossible since the dimension of the problem, i.e. the number of the parameters to be estimated, is $2m$.  This rules out grid based methods to find the MLE. With the increasing number of parameters, the existence of multiple local extrema confounds most optimization methods for finding the global extremum. 

\begin{figure}[htb!]
     \centering
     \begin{subfigure}[b]{0.48\textwidth}
         \centering
         \includegraphics[width=1.1\textwidth]{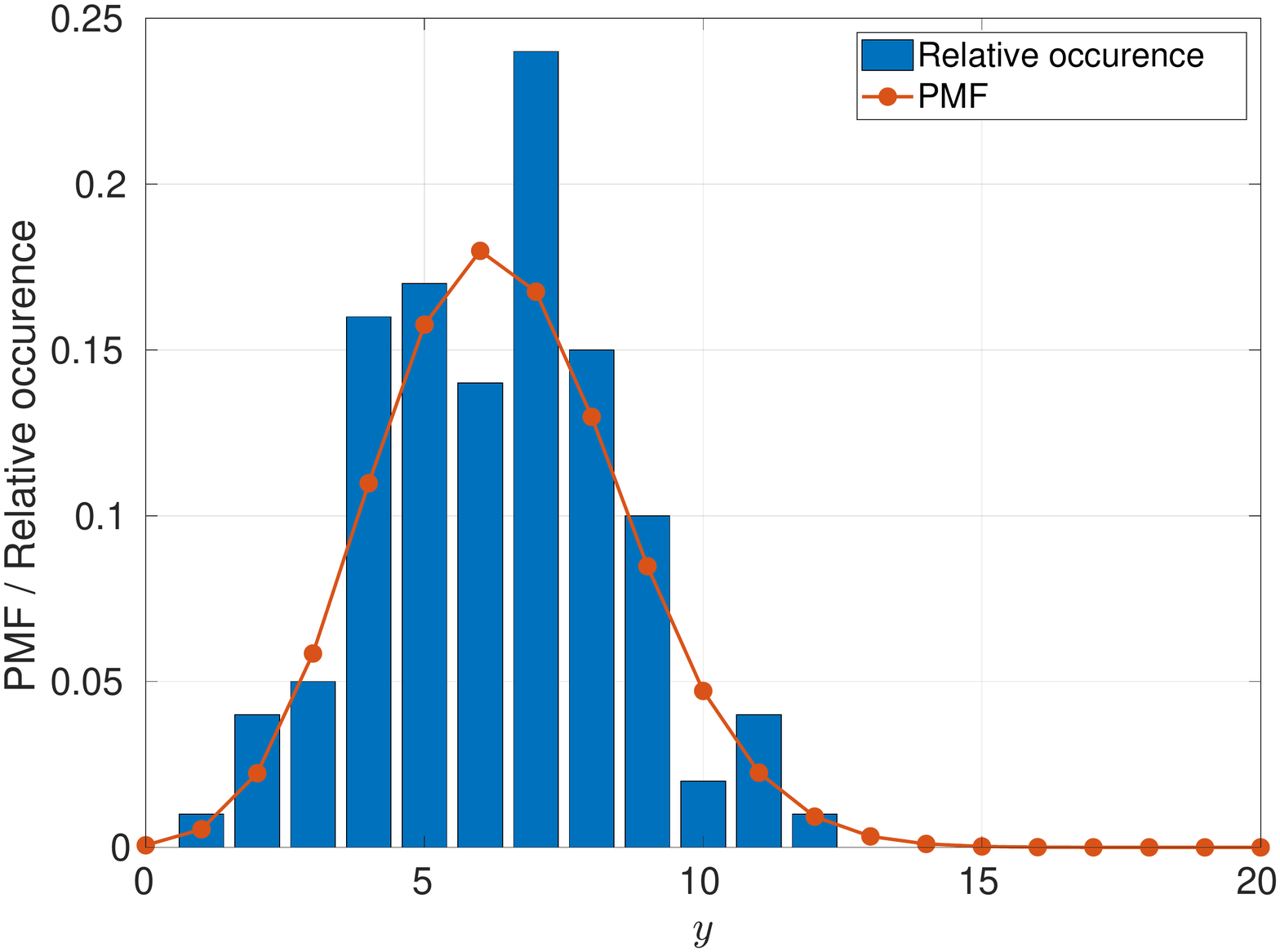}
         \caption{$\nu=100$}
         \label{fig:pdf1}
     \end{subfigure}
     \hspace{-.1cm}
     \begin{subfigure}[b]{0.48\textwidth}
         \centering
         \includegraphics[width=1.1\textwidth]{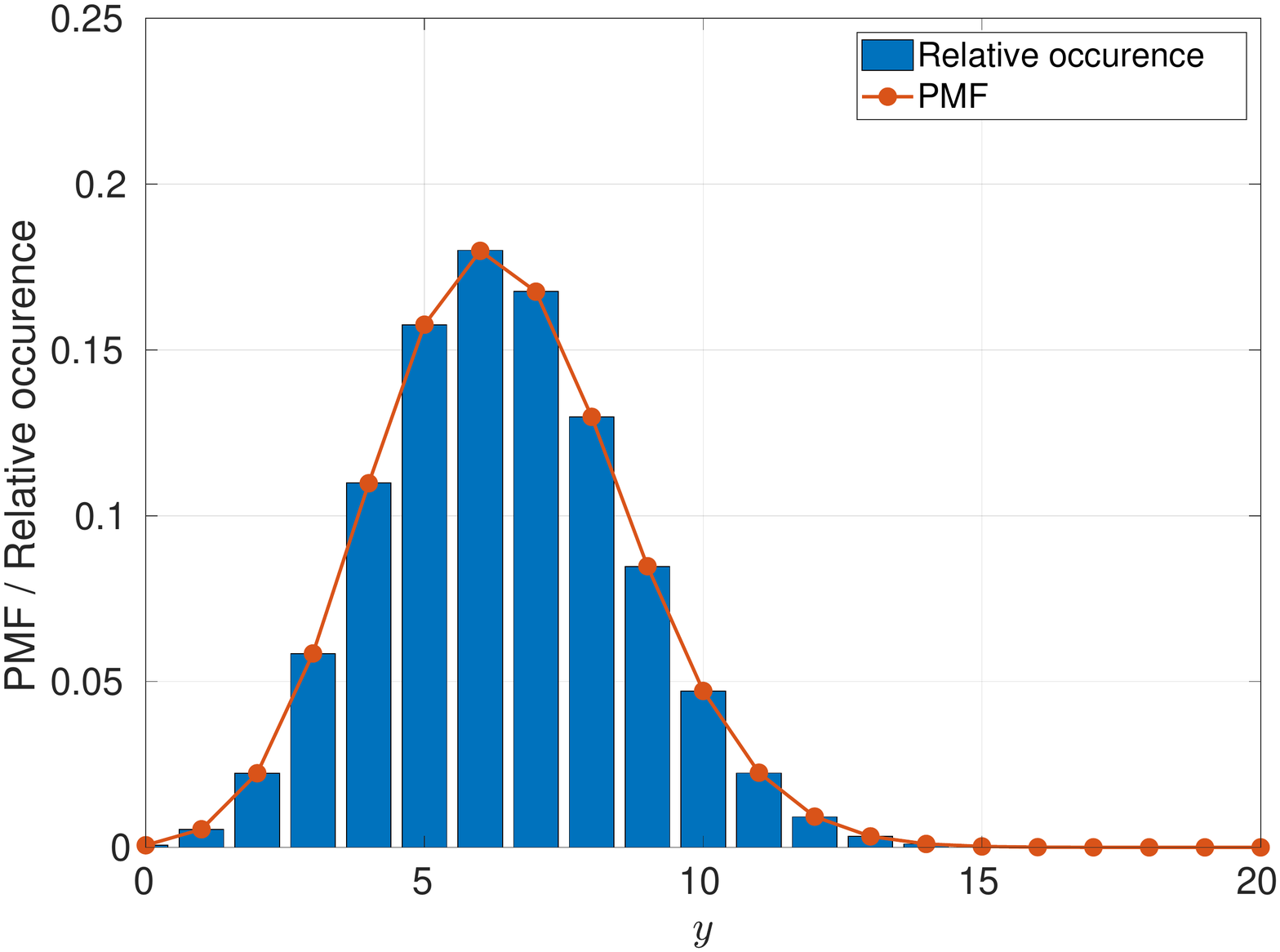}
         \caption{$\nu=1\text{e}7$}
         \label{fig:pdf2}
     \end{subfigure}
     \begin{subfigure}[b]{0.48\textwidth}
         \centering
         \includegraphics[width=1.1\textwidth]{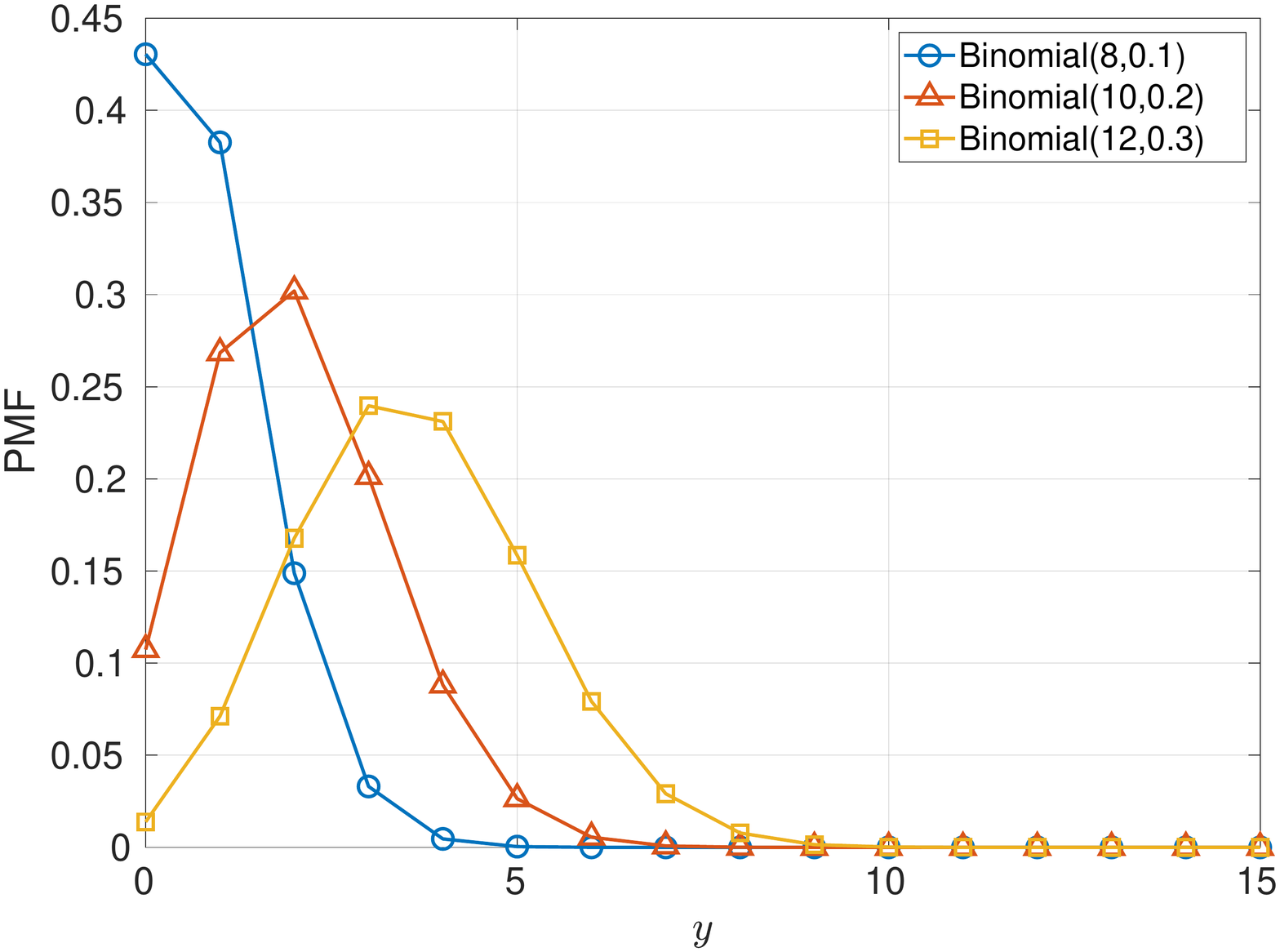}
         \caption{Binomial PMF for each species.}
         \label{fig:pdf3}
     \end{subfigure}
        \caption{The comparison between the relative occurrence obtained  from synthetic measurement from photons with different $\nu$, (a) $\nu = 100$ and (b) $\nu = 1\text{e}7$, where $\bm{\theta}=[8,0.1,10,0.2,12,0.3]$.  (c) shows the Binomial PMF for each species.  As an additional comparison, the Normal approximation for the expected PMF is given. It can be seen that the histogram obtained from the synthetic data converges to the expected distribution of photons with the increasing of the $\nu$.
        }
        \label{fig:pdf}
\end{figure}

To investigate the determination of $\bm{\theta}$, we generated synthetic data using the PMF, Eq.~\eqref{pdf}, with  total number of experiments $\nu$. These synthetic data yield a histogram of events, as illustrated in Fig.~\ref{fig:pdf}. The synthetic data was generated on the basis of $\nu= 100$ experiments, with parameter $\bm{\theta}=[8,0.1,10,0.2,12,0.3]$. Also shown in Fig.~\ref{fig:pdf1}, the expected PMF given by Eq.~\eqref{pdf}, the histogram of synthetic data when $\nu=100$, the Gaussian approximation to Eq.~\eqref{pdf} are given. As the number of experiments increases, the synthetic data should
converge to the expected PMF as shown in Fig.~\ref{fig:pdf2}.

\begin{remark}\label{rmk1}
    In the estimation of the parameters of fluorescent species, one may be more interested in the number of emitters of each species rather than the probabilities, so that, ideally, more consideration should be given to that problem. However, in the estimation of $[M_j,p_j]$, their accuracies are correlated; that is, reduced  accuracy of estimation of $p_j$ will lead to a worse estimation performance for $M_j$. Additionally, because of the integral nature of $M_j$, the estimator $\hat{M}_j$ may be well away from ground truth, corresponding to a small change of  $\hat{p}_j$. Accordingly, we will not  focus on a particular parameter in what follows. 
\end{remark}

\section{Expectation-Maximization algorithm}

The EM algorithm is an effective method for finding the MLE (or local extremum of the likelihood) iteratively by a  simplification of a complicated likelihood function. By carefully choosing the initial guess of the EM, the MLE is approached with high probability. 

In this section, the likelihood function $\ell(C_0\cdots,C_{N}|\bm{\theta})$, i.e. \eqref{likelihood}, is firstly reformulated and simplified to an equivalent problem under the EM framework using the sum-of-log-of-sums method~\cite{hunter2019expansive}. As a result, the $2m$ dimension MLE problem is converted into $m$ independent $2$-dimension optimization problems for which local extrema can be found iteratively, see~\eqref{prob_split}. 

Since the likelihood has many local minima, the initial guess has a significant impact on the final estimate of the EM algorithm, i.e. the EM algorithm may converge to a local critical point with an ``incorrect'' initial guess. It should be noted that EM becomes more sensitive to the initial guess with increasing $m$, as the number of  local  extrema  increases for larger $m$.  Seeding the  EM algorithm with a limited number of random initial guesses typically provides convergence  to the MLE when $m=2$, but the number of initial seeds  becomes unacceptably large when $m> 2$. To overcome this problem, we combined the moment estimator with EM algorithm; that is, the results of  moment estimator are used as the initial guesses. 

\subsection{Reformulating the likelihood function for EM algorithm}

We introduce an intermediate variable $\hat{\bm{\theta}}^{\langle s\rangle}$  that is the estimate of $\bm{\theta}$ at the $s$-th iteration of the EM algorithm. Given  $\hat{\bm{\theta}}^{\langle s\rangle}$, from \eqref{likelihood}, we have

\begin{align}
\log L(\mathbf{y}|\bm{\theta})-\log L(\mathbf{y}|\hat{\bm{\theta}}^{\langle s\rangle})
=&\log\left(\frac{\sum_{\mathbf{y}\in\mathcal{Y}_i} \frac{{f}(\mathbf{y}|\bm{\theta})}{{f}(\mathbf{y}|\hat{\bm{\theta}}^{\langle s\rangle})}{f}(\mathbf{y}|\hat{\bm{\theta}}^{\langle s\rangle})}{L(\mathbf{y}|\hat{\bm{\theta}}^{\langle s\rangle})}\right)\\
\geq&\sum_{\mathbf{y}\in\mathcal{Y}_i} w_{\mathbf{y}}^{\langle s\rangle}\log\left(\frac{f(\mathbf{y}|\bm{\theta})}{f(\mathbf{y}|\hat{\bm{\theta}}^{\langle s\rangle})}\right)\label{jensen}\\
=&\sum_{\mathbf{y}\in\mathcal{Y}_i} w_{\mathbf{y}}^{\langle s\rangle}\log f(\mathbf{y}|\bm{\theta})-\sum_{\mathbf{y}\in\mathcal{Y}_i} w_{\mathbf{y}}^{\langle s\rangle}\log f(\mathbf{y}|\hat{\bm{\theta}}^{\langle s\rangle})\\
=&\tilde{Q}(\bm{\theta}|\hat{\bm{\theta}}^{\langle s\rangle})-\tilde{Q}(\hat{\bm{\theta}}^{\langle s\rangle}|\hat{\bm{\theta}}^{\langle s\rangle})
\end{align}
where \eqref{jensen} follows by Jensen's inequality~\cite{hunter2019expansive}, with $w_{\mathbf{y}}^{\langle s\rangle}=\frac{{f}(\mathbf{y}|\hat{\bm{\theta}}^{\langle s\rangle})}{L(\mathbf{y}|\hat{\bm{\theta}}^{\langle s\rangle})}$ and $\sum_{\mathbf{y}\in\mathcal{Y}_i}w_{\mathbf{y}}^{\langle s\rangle}=1$.

In a similar way, we consider the joint log-likelihood function $\ell(C_0,\cdots,C_{N}|\bm{\theta})$. In this case,  the auxiliary function is
\begin{align}
Q(\bm{\theta}|\hat{\bm{\theta}}^{\langle s\rangle})&=\sum_{i=0}^{N}C_i\sum_{\mathbf{y}\in\mathcal{Y}_i} w_{\mathbf{y}}^{\langle s\rangle}\log f(\mathbf{y}|\bm{\theta})\notag\\
&=\sum_{i=0}^{N}C_i\sum_{\mathbf{y}\in\mathcal{Y}_i} w_{\mathbf{y}}^{\langle s\rangle}\sum_{j=1}^m\log g(y_j|\bm{\theta}_j)
\end{align}
where $f(\mathbf{y}|\bm{\theta})$ is defined in \eqref{likelihood} and $g(y_{j}|\bm{\theta}_j)=\frac{M_j}{(M_j-y_{j})!y_{j}!}p_j^{M_j-y_{j}}(1-p_j)^{y_{j}}$. One can find a local extremum of $\log L(\mathbf{y}|\bm{\theta})$ by maximizing $\tilde{Q}(\bm{\theta}|\hat{\bm{\theta}}^{\langle s\rangle})$ iteratively~\cite{wu1983convergence}. 
The problem then becomes to maximize $Q(\bm{\theta}|\hat{\bm{\theta}}^{\langle s\rangle})$ over $\bm{\theta}$, i.e.
\begin{align}
\hat{\bm{\theta}}^{\langle s+1\rangle}=\left\{\hat{\bm{\theta}}_1^{\langle s+1\rangle},\cdots,\hat{\bm{\theta}}_m^{\langle s+1\rangle}\right\}=\arg\max_{\mathbf{y}\in\varprod_{j=1}^m \left([0,1]\times\mathbb{Z}^+\right)}Q(\bm{\theta}|\hat{\bm{\theta}}^{\langle s\rangle})\label{MLE2}
\end{align}

Since, $\log g(y_j|\bm{\theta}_j)$, $j=1,2,\cdots,m$, are independent with respect to $\bm{\theta}_{j^-}$, where $j^-=\{1,\cdots,m\}\setminus\{j\}$,  \eqref{MLE2} can be rewritten as
\begin{align}
\hat{\bm{\theta}}^{\langle s+1\rangle}=&\Bigg\{\hat{\bm{\theta}}_1^{\langle s+1\rangle}=\arg\max_{\bm{\theta}_1\in [0,1]\times\mathbb{Z}^+}\sum_{i=0}^{N}C_i\sum_{\mathbf{y}\in\mathcal{Y}_i} w_{\mathbf{y}}^{\langle s\rangle}\log g(y_1|\bm{\theta}_1),\;\cdots,\notag\\
&\;\;\hat{\bm{\theta}}_m^{\langle s+1\rangle}=\arg\max_{\bm{\theta}_m\in [0,1]\times\mathbb{Z}^+}\sum_{i=0}^{N}C_i\sum_{\mathbf{y}\in\mathcal{Y}_i} w_{\mathbf{y}}^{\langle s\rangle}\log g(y_m|\bm{\theta}_m)\Bigg\}\notag\\
=&\left\{\hat{\bm{\theta}}_1^{\langle s+1\rangle}=\arg\max_{\bm{\theta}_1\in [0,1]\times\mathbb{Z}^+}Q_1\left(\bm{\theta}_1|\hat{\bm{\theta}}^{\langle s\rangle}\right),\cdots,\hat{\bm{\theta}}_m^{\langle s+1\rangle}=\arg\max_{\bm{\theta}_m\in [0,1]\times\mathbb{Z}^+}Q_m\left(\bm{\theta}_m|\hat{\bm{\theta}}^{\langle s\rangle}\right)\right\}\label{prob_split}
\end{align}

For  $j\in\{1,\cdots,m\}$, we have
\begin{align}
&\hat{\bm{\theta}}_j^{\langle s+1\rangle}=\arg\max_{\bm{\theta}_j\in [0,1]\times\mathbb{Z}^+}Q_j\left(\bm{\theta}_j|\hat{\bm{\theta}}^{\langle s\rangle}\right)\notag\\
\Longrightarrow\;\;\;&\frac{\partial}{\partial p_j}Q_j\left(\bm{\theta}_j|\hat{\bm{\theta}}^{\langle s\rangle}\right)=0\notag\\
\Longrightarrow\;\;\;&\hat{p}_j^{\langle s+1\rangle}=\frac{\sum_{i=0}^{N}C_i\sum_{\mathbf{y}\in\mathcal{Y}_i}w_{\mathbf{y}}^{\langle s\rangle}y_j}{\hat{M}^{\langle s\rangle}_j\nu}\label{estimator_p}
\end{align}
Substituting $\hat{p}^{\langle s+1\rangle}_j$ into $j$-th term of \eqref{prob_split}, we obtain  
\begin{align}
\hat{\bm{\theta}}_j^{\langle s+1\rangle}=\left\{\hat{M}_j^{\langle s+1\rangle}=\arg\max_{{M}_j\in\mathbb{Z}^+}Q_j\left({M}_j|\hat{p}^{\langle s+1\rangle}_1,\hat{\bm{\theta}}^{\langle s\rangle}\right),\;\;\hat{p}_j^{\langle s+1\rangle}\right\},\;\;\;j=1,\cdots,m \label{prob_split2}
\end{align}

Now, the EM algorithm can be implemented using an initial guess $\hat{\bm{\theta}}_j^{\langle 0\rangle}$ for $j=1,\cdots,m$ and the (local) estimate can be obtained until $\hat{\bm{\theta}}_j^{\langle s\rangle}$ converges. The structure of the EM algorithm is listed in Algorithm \ref{alg:one}.

\begin{algorithm}[htb!]
\caption{The Structure of the EM algorithm to Estimate $\bm{\theta}$}\label{alg:one}
\KwData{$C_0,\cdots,C_{N}$}
\KwResult{$\hat{\bm{\theta}}\gets[\hat{M}_{1},\hat{p}_{1},\cdots,\hat{M}_{m},\cdots,\hat{p}_{m}]$}
$s \gets 0$\;
Choose initial guesses $[\hat{M}^{\langle s\rangle}_1,\cdots,\hat{M}^{\langle s\rangle}_m]$ and $[\hat{p}^{\langle s\rangle}_1,\cdots,\hat{p}^{\langle s\rangle}_m]$

    \Repeat{converge}{
     $s \gets s+1$\;
    Calculate $\hat{p}^{\langle s\rangle}_{j}$ using \eqref{estimator_p} and \eqref{prob_split2} for $j=1,\cdots,m$\;
}
$\hat{\bm{\theta}}\gets[\hat{M}^{\langle s\rangle}_1,{p}^{\langle s\rangle}_1,\cdots,\hat{M}^{\langle s\rangle}_m,{p}^{\langle s\rangle}_m]$\;

\end{algorithm}

\subsection{Choice of the initial guess for the EM algorithm}

In the EM algorithm, the log likelihood is guaranteed to increase at each EM iteration, and it converges to a maximum of the likelihood under mild conditions~\cite{wu1983convergence}. However, it is unnecessary to be the global optimizer.

As an ``always improving'' algorithm~\cite{dempster1977maximum}, EM is, of course, sensitive to the initial guess of $\bm{\theta}$, i.e., $\hat{\bm{\theta}}^{\langle 0\rangle}$ when the likelihood function contains multiple critical points. We observe that the number of local critical points increases dramatically with the increasing number of  species.

\subsubsection{Choosing initial guesses for $[M_1,\cdots,M_m]$}
When $m>1$, \eqref{prob_split2} may converge to a local minimum that is not the MLE. To relieve this problem, a search procedure can be adopted into the EM algorithm. In \eqref{prob_split2}, it can be observed that $p^{\langle s+1\rangle}_j$ is updated at each iteration step using $\hat{M}_j^{\langle s\rangle}$. By taking into account that  $p^{\langle s+1\rangle}_j$ has a better chance to converge to $p_j$ when $\hat{M}_j^{\langle s\rangle}$ converges to $M_j$, one can fix $\hat{M}_j^{\langle s\rangle}={M}_j$, $\forall s$ and $j=1,\cdots,m$, in \eqref{prob_split2} and then find the optimized $\hat{p}^{\langle s\rangle}_j$ given $[{M}_1,\cdots,{M}_m]$ iteratively. 

In practice, $[{M}_1,\cdots,{M}_m]$ is unknown and to be estimated. However, since $M_1,\, \ldots,\, M_m$ are (bounded) integers so that their possible values are finite and listable by enumeration. Suppose that $M_j\leq M$, $\forall j$ and $M\in\mathbb{Z}^+$, then a set, $\mathcal{M}=\{\mathcal{M}_1,\cdots,\mathcal{M}_{L}\}$, containing all possible combinations for $[{M}_1,\cdots,{M}_m]$ can be constructed by $m$-combinations from the integer set $\{1,2,\cdots,M\}$ without repetition, order does not matter and it can be verified that $L=\binom{M}{n}$. In other words, $\mathcal{M}_l$, $l=1,\cdots,L$, corresponds to a possible solution (combination) to $[{M}_1,\cdots,{M}_m]$. The EM algorithm is, therefore, implemented $L$ times. At  the $l$-th implementation, the guess $[\hat{M}_1^{\langle 0\rangle},\cdots,\hat{M}_m^{\langle 0\rangle}]$ is chosen to be $\mathcal{M}_l$ and fixed for all $s$. For simplicity, we denote $[\hat{M}_{m,l},\cdots,\hat{M}_{m,l}]$ as the guess of $[{M}_1,\cdots,{M}_m]$ at $l$-th implementation of EM algorithm. 

\subsubsection{Choice of  initial guesses for $[p_1,\cdots,p_m]$}

To obtain initial guesses for the estimator of $[p_1,\cdots,p_m]$, we use the estimates from the moment estimator. By using the data, it is straightforward to calculate the sample  mean, $\hat{\mu}_1=\sum_{i=0}^{N}\frac{C_i}{\nu}i$, and $k$-th sample  central moments, $\hat{\mu}_k=\sum_{i=0}^{N}\frac{C_i}{\nu}(i-\hat{\mu}_1)^k$ for $k>1$. The associated population  mean and central moments are $\mu_1=\mathbf{E}[x]$ and $\mu_k=\mathbf{E}[(x-\mu_1)^k]$ for $k>1$.
Then the moment estimator, $\hat{\bm{\theta}}^{mom}$, can be obtained by solving $\hat{\mu}_i={\mu}_i$ for $i=1,\cdots,m$, and this can be used to seed the EM algorithm: $\hat{\bm{\theta}}^{\langle 0\rangle}=\hat{\bm{\theta}}^{mom}$. Since we may find  multiple $\hat{\bm{\theta}}^{mom}$ from the moment estimator,the EM can be run in parallel with different initial guesses.

As an example, the moments $\mu_i$ of the sum of $m=4$ Binomial distributed random variables are, \cite{butler2017distribution},
\begin{align}
\begin{cases}
\displaystyle \mu_1=\sum_{j=1}^mM_jp_j\\
\displaystyle \mu_2=\sum_{j=1}^{m}\left(1-p_j\right)M_jp_j\\
\displaystyle \mu_3=\sum_{j=1}^{m}\left(1-p_j\right)\left(1-2p_j\right)M_jp_j\\
\displaystyle \mu_4= \sum_{j=1}^{m}M_j p_j\left(1-p_j\right)\left(1+\left(3 M_j-6\right) \left(1-p_j\right)p_j\right),
\end{cases}
\label{equ_initial1}
\end{align}
which can be simplified to
\begin{align}
\begin{cases}
\displaystyle \sum_{j=1}^mM_jp_j={\mu}_1\\
\displaystyle \sum_{j=1}^mM_jp_j^2={\mu}_1-{\mu}_2\\
\displaystyle \sum_{j=1}^mM_jp_j^3=\frac{1}{2} \left(2 {\mu}_1 - 3 {\mu}_2 + {\mu}_3\right)\\
\displaystyle \sum_{j=1}^mM_jp_j^4=\frac{1}{6}\left(6{\mu}_1 - 11{\mu}_2 + 3{\mu}_2^2 + 6{\mu}_3 - {\mu}_4\right)\\
\end{cases}\label{equ_initial2}
\end{align}

Applying  the method of moments, we replace $[{M}_1,\cdots,{M}_m]$ by  $[\hat{M}_{1,l},\cdots,\hat{M}_{m,l}]$ and $[{\mu}_1,\cdots,{\mu}_m]$ by  $[\hat{\mu}_{1,l},\cdots,\hat{\mu}_{m,l}]$ in \eqref{equ_initial2} and find the real solutions  $[\hat{p}_{1,l},\cdots,\hat{p}_{m,l}]$ to provide  the initial guess of $[p_1,\cdots,p_m]$ at the $l$-th implementation of EM algorithm. After obtaining multiple candidate estimates with different initial guesses, the estimated MLE among these candidates is the one with the largest  value of the likelihood function \eqref{likelihood}. The algorithm is summarized in Algorithm \ref{alg:two}.

\begin{algorithm}[htb!]
\caption{The algorithm to estimate $\bm{\theta}$ combining searching strategy}\label{alg:two}
\KwData{$C_0,\cdots,C_{N}$}
\KwResult{$\hat{\bm{\theta}}\gets[\hat{M}_{1},\hat{p}_{1},\cdots,\hat{M}_{m},\cdots,\hat{p}_{m}]$}
$\{\mathcal{M}_1,\cdots,\mathcal{M}_L\} \gets \text{nchoosek}(N,m)$\;
$k\gets 1$\; 
\For{$l=1:L$}{
$sol = \text{solve}\left(\sum_{j=1}^m\hat{M}_{j,l}p_j^i=\hat{\mu}_i,\;\;\;i=1,\cdots,m\right)$\;
  \If{$sol$ is real}{
  $s \gets 0$\;
  $\hat{p}^{\langle s\rangle}_{1,l},\cdots,\hat{p}^{\langle s\rangle}_{m,l}\gets sol$\;
    \Repeat{converge}{
     $s \gets s+1$\;
    Calculating $\hat{p}^{\langle s\rangle}_{j,l}$ using \eqref{prob_split2} by replacing $\hat{M}^{\langle s\rangle}_j$ with $\hat{M}_{j,l}$\;
   }
     $\hat{\bm{\theta}}_k=[\hat{M}_{1,l},\hat{p}^{\langle s\rangle}_{1,l},\cdots,\hat{M}_{m,l},\hat{p}^{\langle s\rangle}_{m,l}]$\;
  $e_k=\ell(\hat{\bm{\theta}}_l|C_0,\cdots,C_{N})$\;
  $k\gets k+1$ \;
  }
  $I\gets\displaystyle{\arg\max_{i}}\;\{e_i\}$\;
$\hat{\bm{\theta}}\gets\hat{\bm{\theta}}_I$\;
}
\end{algorithm}


\section{Cr\'amer-Rao lower bound}
In this section, we calculate the Fisher Information Matrix  (FIM) and then the
Cram\'er-Rao lower bound (CRLB)~\cite{Ly2017jmp}. The CRLB provides a lower bound for the variance of an unbiased estimator. The underlying likelihood function \eqref{likelihood} contains continuous as well as discrete components so that the conventional method to derive CRLB may not be applicable. A Cram\'er-Rao type bound for discrete likelihood function is proposed in \cite{Nishiyama2019arxiv}. However, it cannot handle the distribution containing discrete and continuous components. In this paper, following the calculation  in~\cite{Li2022oe}, an approximated CRLB is derived by approximating the discrete component $x!$ by a continuous function $x\Gamma(x)$. On the other hand, the MLE is typically asymptotically unbiased under mild conditions~\cite{lehmann2006theory}. From the simulation  given in Section \ref{sec_simulation}, the proposed MLE asymptotically approaches the derived approximated CRLB.

 For the parameter $\bm{\theta}= [\theta_1,\theta_2,\cdots,\theta_{2m-1},\theta_{2m}]=[M_1,p_1,\cdots,M_m,p_m]$, we see that the $(k,l)$-th element of the FIM for \eqref{pdf}, $\mathbf{I}(\bm{\theta})_{k,l}$ is
\begin{align}
\mathbf{I}(\bm{\theta})_{k,l}&=\sum_{i=0}^M\sum_{ \mathbf{y}\in\mathcal{Y}_i}\left\{\left(\frac{ \partial f(\mathbf{y}|\bm{\theta})}{ \partial \theta_{k}}\frac{ \partial f(\mathbf{y}|\bm{\theta})}{ \partial \theta_{l}}\right)\frac{1}{f(\mathbf{y}|\bm{\theta})}\right\}\label{FIM}
\end{align}
where $f(\mathbf{y}|\bm{\theta})=\prod_{j=1}^m pr_j(y_j|M_j,p_j)$. 


By using  $x!=x\Gamma(x)$ and $\left[x\Gamma(x)\right]'=\Gamma(x)+x\Gamma(x)\psi(x)$, where $\Gamma(\cdot)$ is the Gamma function and $\psi(\cdot)$ is the digamma function, we have 
\begin{align}
&\frac{ \partial pr_j(y_j|M_j,p_j)}{ \partial M_j}\notag\\
=&\frac{\Gamma (M_j) p^{y_j} (1-p_j)^{M_j-y_j} }{y_j! (M_j-y_j)^2 \Gamma (M_j-y_j)}\left\{M_j (y_j-M_j) \left[\psi(M_j-y_j)-\psi(M_j)-\log (1-p_j)\right]-y_j\right\}\label{FIM_M}
\end{align}
and 
\begin{align}
\frac{ \partial f(y_j|M_j,p_j)}{ \partial p_j}=-\frac{M_j! p_j^{y_j-1} (1-p_j)^{M_j-y_j-1} (M_j p_j-y_j)}{y_j! (M_j-y_j)!}\label{FIM_p}
\end{align}

As a result, by applying chain rule, we are able to calculate $\frac{ \partial f(\mathbf{y}|\bm{\theta})}{ \partial \theta_{k}}$ using \eqref{FIM_M} and \eqref{FIM_p}. Then the FIM, $\mathbf{I}(\bm{\theta})$, can  now be calculated by inserting \eqref{FIM_M} and \eqref{FIM_p} into \eqref{FIM}. The  CRLB is just $\mathbf{C}(\bm{\theta})=\mathbf{I}(\bm{\theta})^{-1}$. For  $\nu$ experiments, the CRLB at ground truth $\bm{\theta}_0$ is
$$\mathbf{C}_{\nu}(\bm{\theta}_0)=\frac{1}{\nu}\mathbf{C}(\bm{\theta})\Big|_{\bm{\theta}=\bm{\theta}_0}.$$

\section{Simulation}\label{sec_simulation}
Here, the algorithm is evaluated via Monte Carlo (MC) simulations. The metrics used  to evaluate performance are
\begin{align}
    \text{RMSE}(\hat{X}_{1:nMC,j},X_j) =& \sqrt{\sum_{i=1}^{nMC}\frac{\left(\hat{X}_{i,j}-X_{j}\right)^2}{nMC}}\label{rmse}\\
    \text{aMAPE}(\hat{X}_{1:nMC,1:4},X_{1:4}) =&\frac{1}{m}\sum_{j=1}^m \left(\sum_{i=1}^{nMC}\left|\frac{\hat{X}_{i,j}-X_j}{X_j}\right|\right)\label{amape}
\end{align}
where $X\in\{M,P\}$, $\hat{X}_{i,j}$ is the estimate of $X_j$ at the  $i$-th MC simulation, $nMC$ is the number of MC simulations, and $\text{RMSE}(\cdot)$ is the Root Mean Square Error and $\text{aMAPE}(\cdot)$ is the averaged Mean Absolute Percentage Error. 

To explore the convergence of our algorithm, in the following subsections we illustrate the performance for the one, two, and three distinct species.

\subsection{One species}
The case of determining the number and detection probability for one species is highly analogous to the case that we presented in ref.~\cite{Li2022oe}.  We concentrate on the small number regime as this is more pertinent for the case that we are concentrating on with few emitters.

\begin{figure}[htb!]
     \centering
     \begin{subfigure}[b]{0.48\textwidth}
         \centering
         \includegraphics[width=1.1\textwidth]{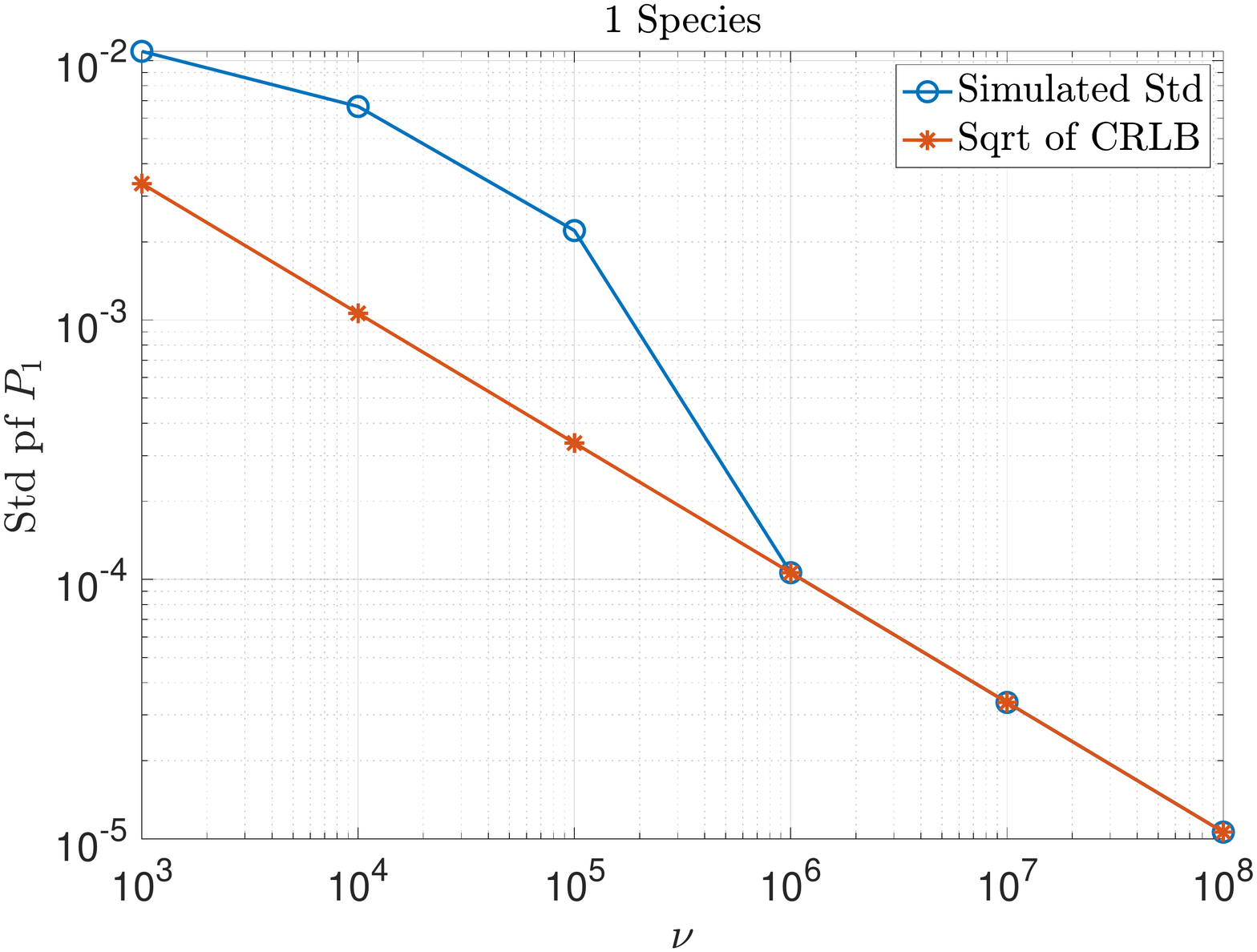}
         \caption{Comparison of $\sqrt{\text{CRLB}}$ and Std of $p_1$.}
         \label{fig:1s_p}
     \end{subfigure}
     \hspace{0cm}
     \begin{subfigure}[b]{0.48\textwidth}
         \centering
         \includegraphics[width=1.1\textwidth]{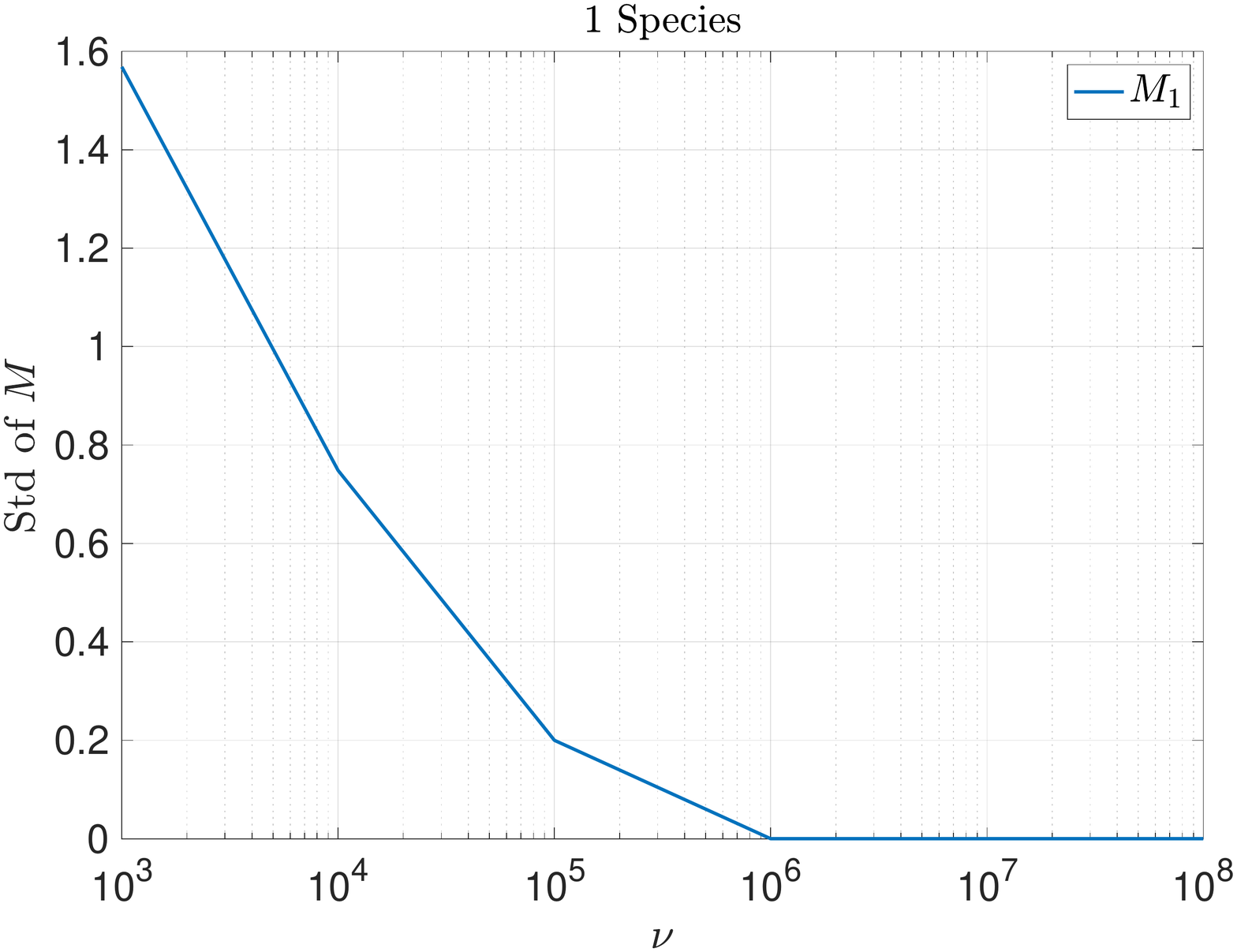}
         \caption{The sampled standard deviation of $M_1$.}
         \label{fig:1s_M}
     \end{subfigure}
        \caption{Given different number of experiments $\nu$, the comparison between the $\sqrt{\text{CRLB}}$ and sampled standard deviation when the number of species is $m=1$.}
        \label{fig:1s}
\end{figure}

Here we assume that the true parameters are $[M_1,p_1]=[8,0.1]$.  Fig.~\ref{fig:1s} shows the convergence of the determination of the number of emitters and detection probability.

\begin{figure}[htb!]
     \centering
         \centering
         \includegraphics[width=0.8\textwidth]{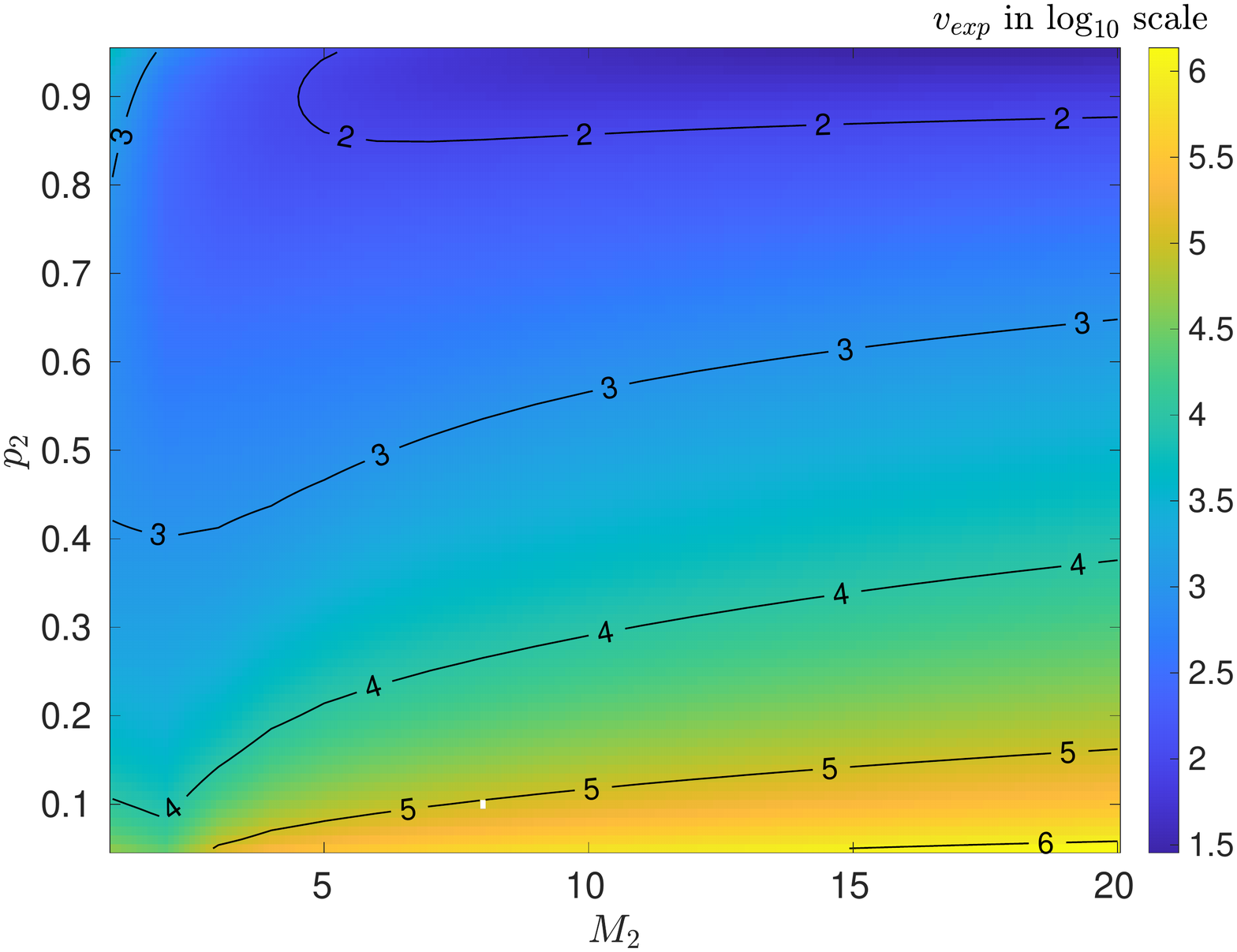}
         \caption{The required number of experiments, $\nu_{exp}$, to attain $\text{CRLB}(M_1)/M_1=1\%$ while $M_1$ varies from $1$ to $20$ and $p_1$ from $0.05$ to $0.95$. Please note that $\nu$ is plotted in $\log_{10}$ scale.}
        \label{fig:nu_CRLB_1}
\end{figure}

Additionally, in Fig.~\ref{fig:nu_CRLB_1}, the (expected) required number of experiments, $\nu_{exp}$, to attain a given estimation performance, $\frac{\text{CRLB}(M_1)}{M_1}=1\%$, with varied $M_1=1,\cdots,20$ and $p_1\in[0.05,0.95]$ is plotted. It can be seen that, to achieve the fixed performance, the small value of $M_1$ or large value of $p_1$ requires less number of experiments, which gives some insight in how the values of parameters relate to the estimation performance.

\subsection{Two species}

\begin{figure}[htb!]
     \centering
     \begin{subfigure}[b]{0.48\textwidth}
         \centering
         \includegraphics[width=1.1\textwidth]{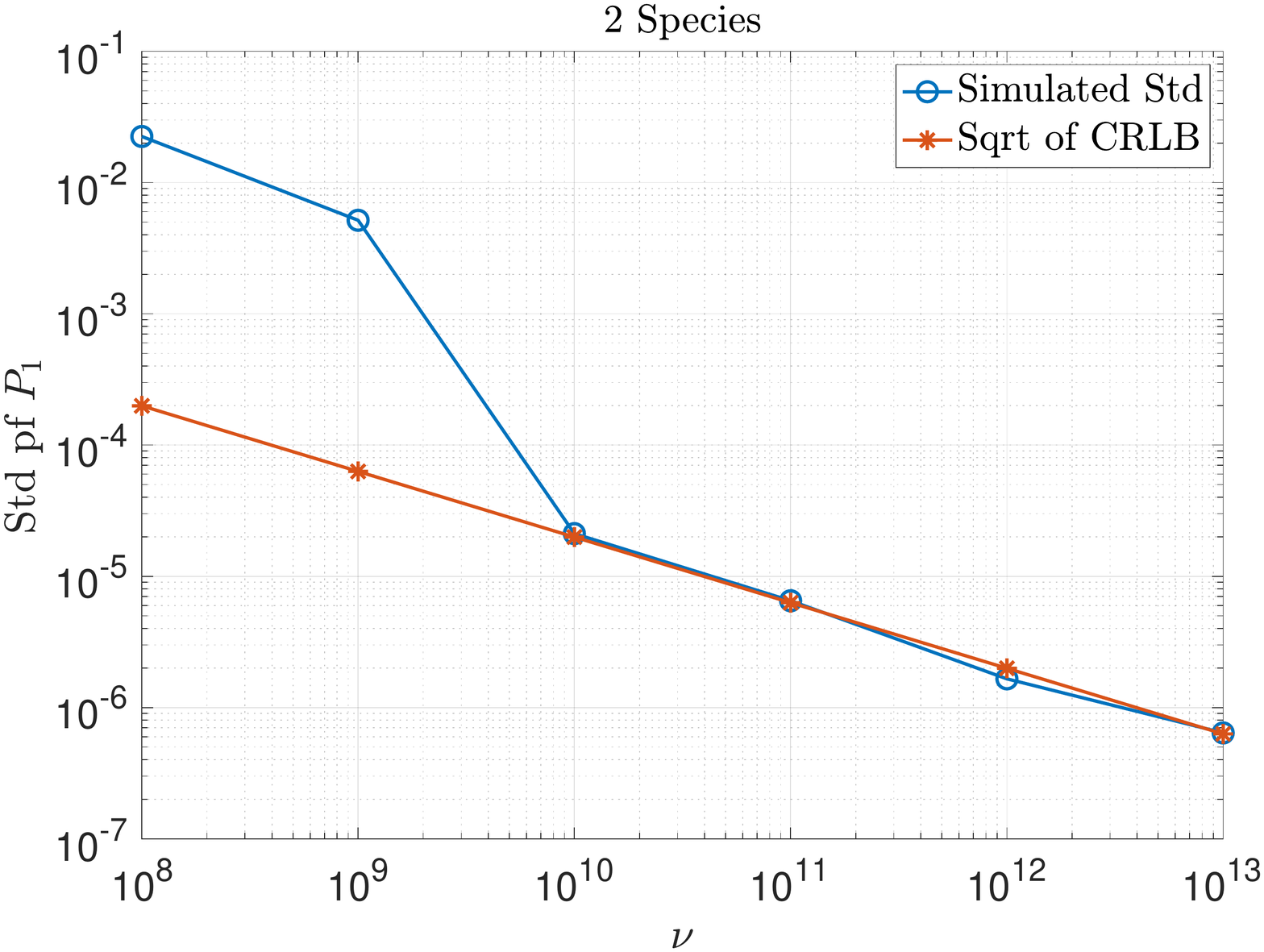}
         \caption{Comparison of $\sqrt{\text{CRLB}}$ and Std of $p_1$.}
         \label{fig:2s_p}
     \end{subfigure}
     \hspace{0cm}
     \begin{subfigure}[b]{0.48\textwidth}
         \centering
         \includegraphics[width=1.1\textwidth]{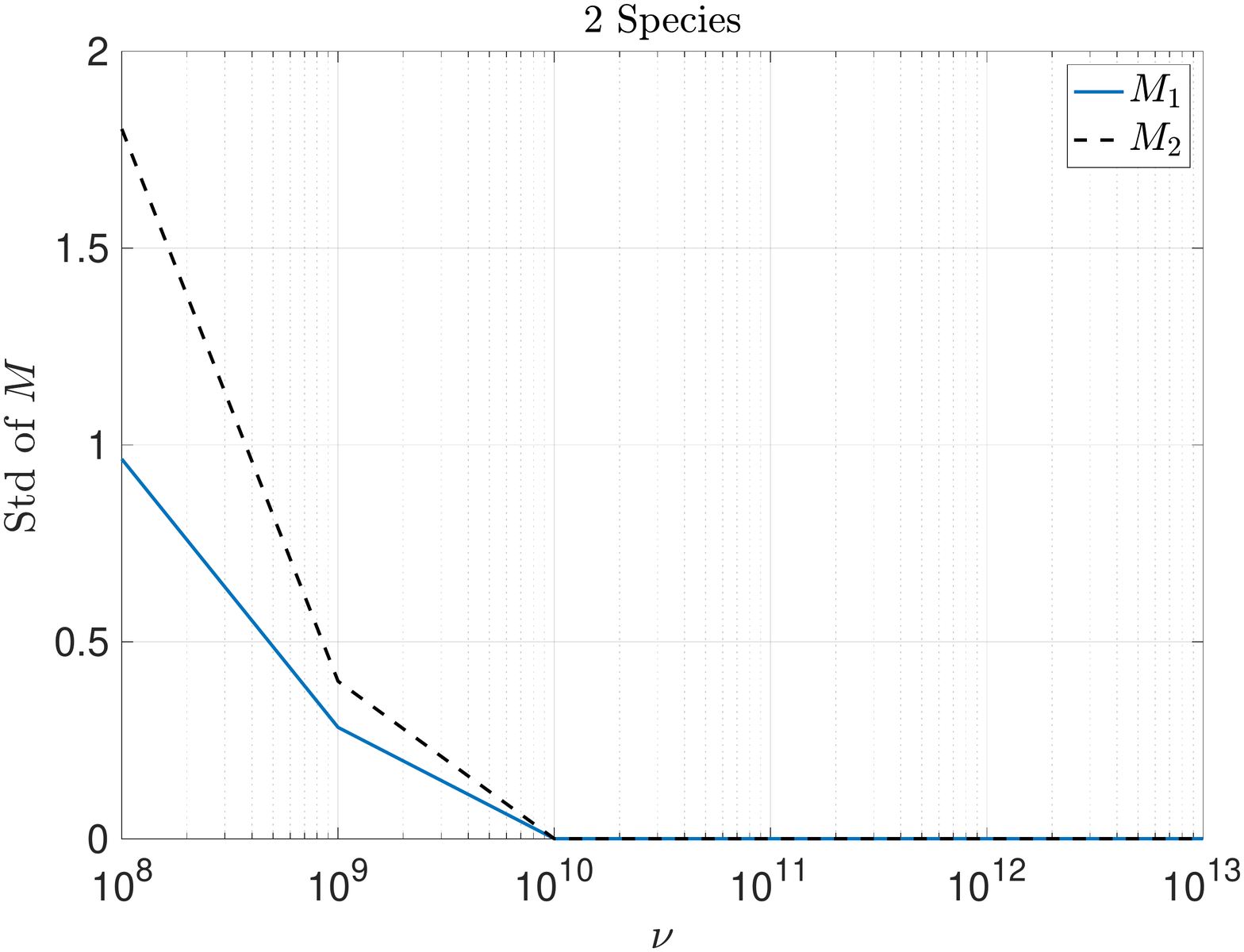}
         \caption{The sampled standard deviation of $M_1$.}
         \label{fig:2s_M}
     \end{subfigure}
        \caption{Given different value of $\nu$, the comparison between the $\sqrt{\text{CRLB}}$ and sampled standard deviation when the number of species is $j=2$.}
        \label{fig:2s}
\end{figure}

In the case of two species, assume that $[M_1,p_1]=[8,0.1]$ and $[M_2,p_2]=[10,0.2]$. The number of experiments, $\nu$, is  set to  different numbers, as  shown in the figures. For each $\nu$, the Monte Carlo simulation is implemented $100$ times. The comparison of the simulated standard deviation and the computed squared root of  the CRLB of $p_1$ for all $\nu$ is shown in Fig.~\ref{fig:2s_p}, while the comparison of the estimates and the standard deviation of the estimated $[M_1,M_2]$ is shown in Fig.~\ref{fig:2s_M} for the  different species. 

\begin{figure}[htb!]
     \centering
         \includegraphics[width=0.8\textwidth]{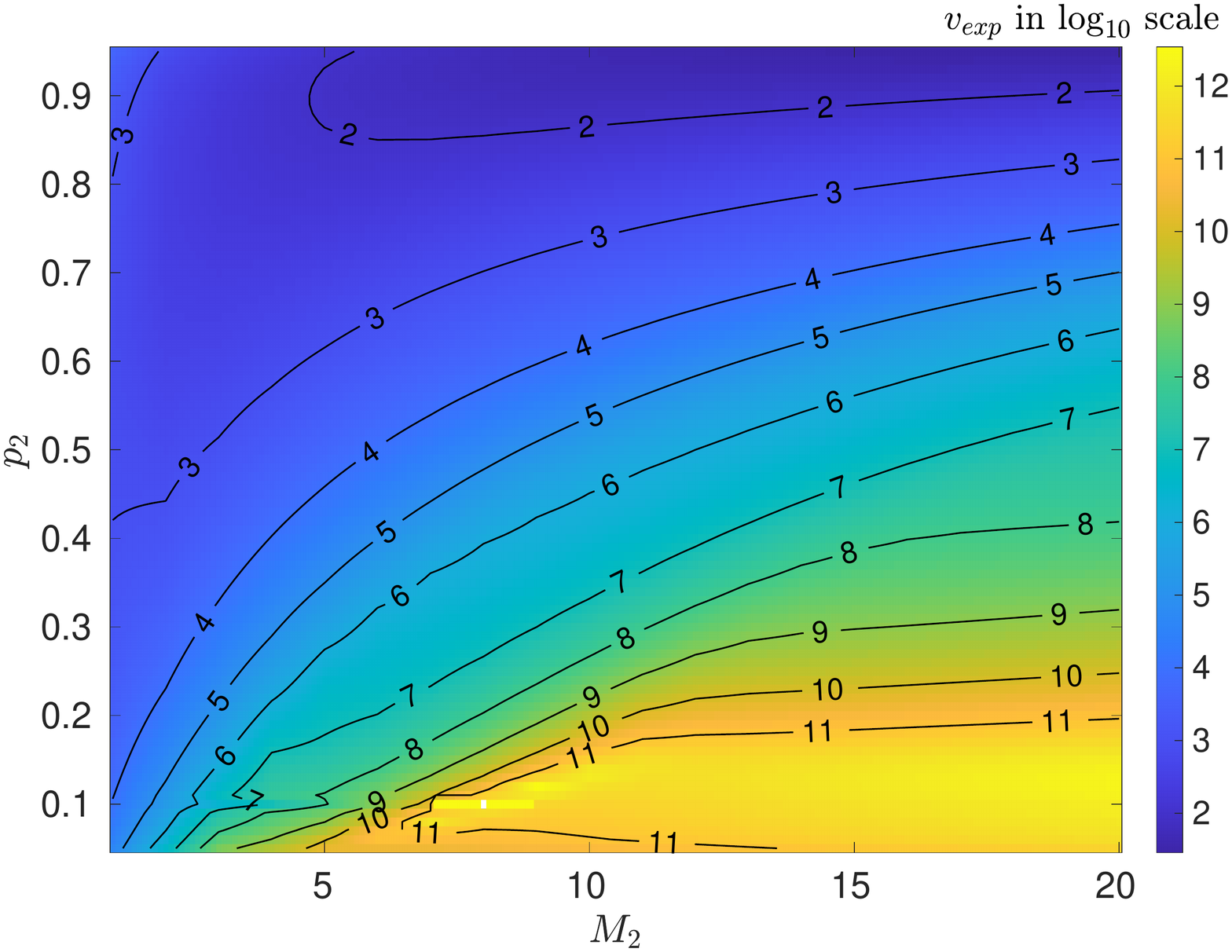}
         \caption{The required number of experiments, $\nu_{exp}$, to attain $\text{CRLB}(M_2)/M_2=1\%$ while $[M_1,p_1]=[8,0.1]$ as well as $M_2$ varies from $1$ to $20$ and $p_2$ from $0.05$ to $0.95$. Please note that $\nu$ is plotted in $\log_{10}$ scale and the white pixel corresponds to the point that $M_1=M_2=8$ and $p_1=p_2=0.1$ so that the $\text{FIM}$ is singular (the CRLB does not exist).}
        \label{fig:nu_CRLB_2}
\end{figure}

Analogue to Fig.~\ref{fig:nu_CRLB_1} for one species, the (expected) required number of experiments, $\nu_{exp}$, to attain $\frac{\text{CRLB}(M_2)}{M_2}=1\%$ with fixed $[M_1,p_1]=[8,0.1]$ and varied $M_2=1,\cdots,20$ and $p_2\in[0.05,0.95]$ is plotted in Fig.~\ref{fig:nu_CRLB_2}. The similar trend can be observed where the small value of $M_2$ and large value of $p_2$ lead to the less $\nu_{exp}$.

\subsection{Three species}
\begin{figure}[htb!]
     \centering
     \begin{subfigure}[b]{0.48\textwidth}
         \centering
         \includegraphics[width=1.1\textwidth]{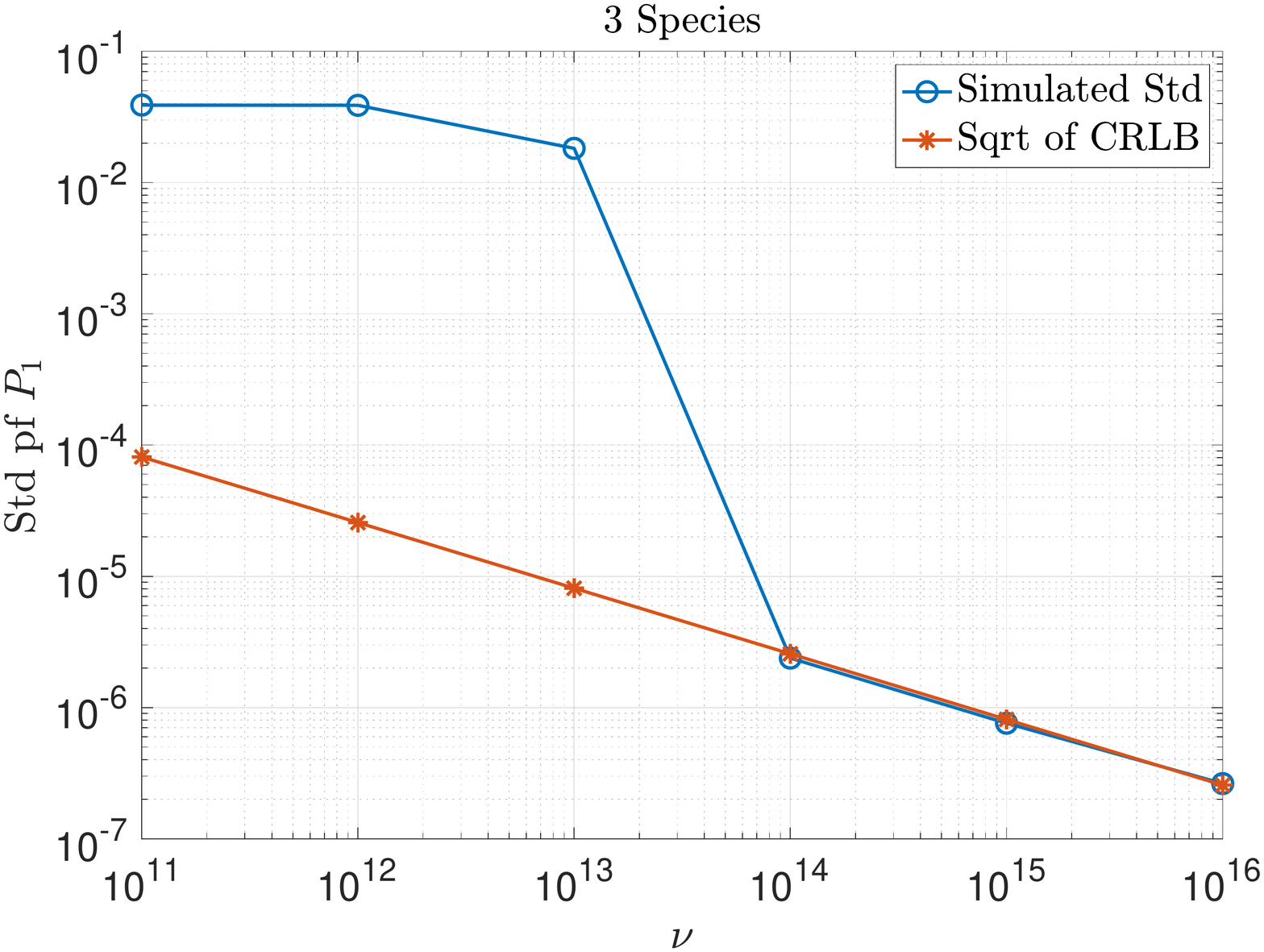}
         \caption{Comparison of $\sqrt{\text{CRLB}}$ and Std of $p_1$.}
         \label{fig:3s_p}
     \end{subfigure}
     \hspace{0cm}
     \begin{subfigure}[b]{0.48\textwidth}
         \centering
         \includegraphics[width=1.1\textwidth]{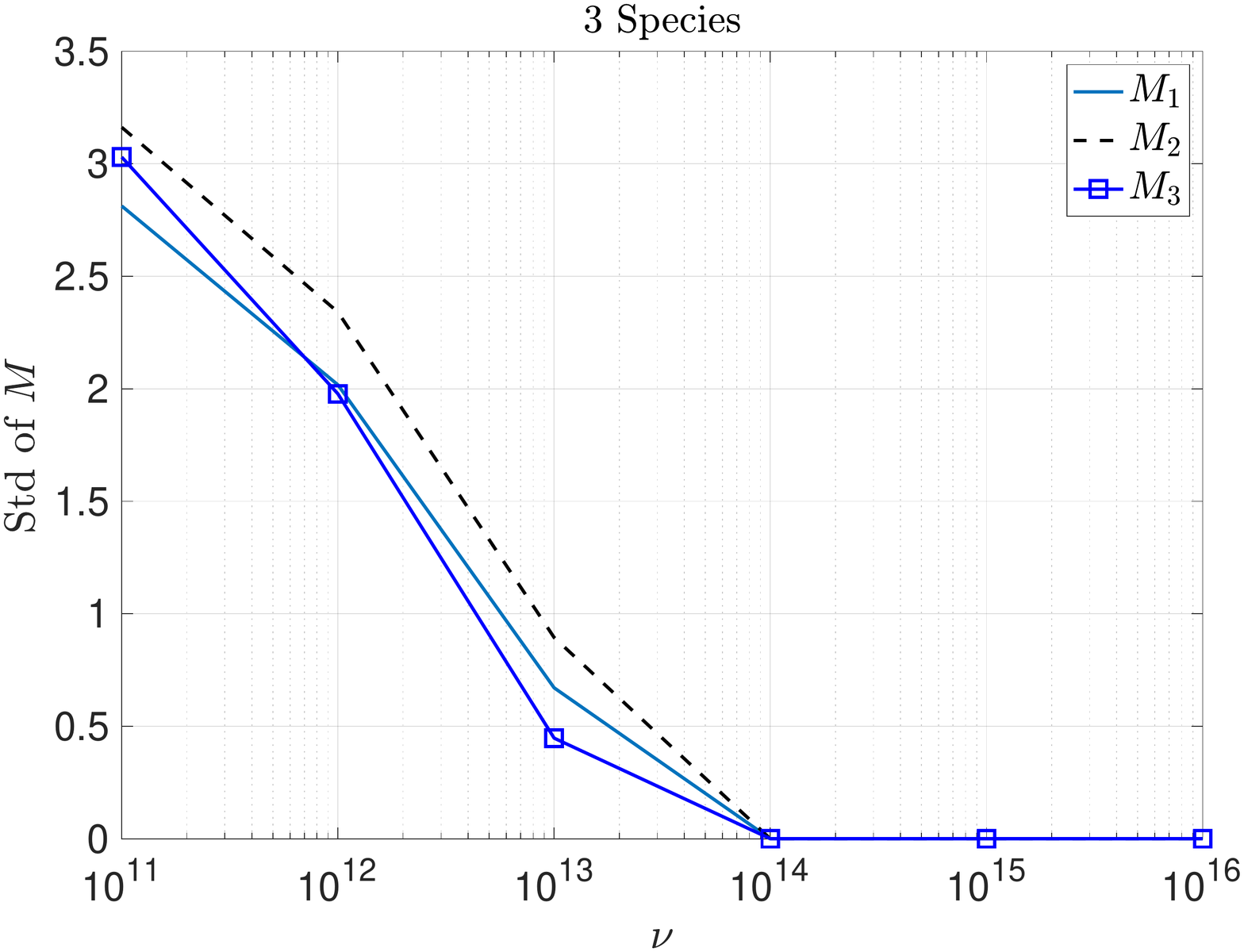}
         \caption{The sampled standard deviation of $M_1$.}
         \label{fig:3s_M}
     \end{subfigure}
        \caption{Given different value of $\nu$, the comparison between the $\sqrt{\text{CRLB}}$ and sampled standard deviation when the number of species is $m=2$.}
        \label{fig:3s}
\end{figure}

In the case of two species, assume that $[M_1,p_1]=[8,0.1]$, $[M_2,p_2]=[10,0.2]$ and $[M_2,p_2]=[12,0.3]$. The number of experiments, $\nu$, is  set to  different numbers, as  shown in the figures. For each $\nu$, the Monte Carlo simulation is implemented $100$ times. The comparison of the simulated standard deviation and the computed squared root of  the CRLB of $p_1$ for all $\nu$ is shown in Fig.~\ref{fig:3s}, while the comparison of the estimates and the standard deviation of the estimated $[M_1,M_2,M_3]$ is shown in Fig.~\ref{fig:3s} for the  different species. 

\begin{figure}[htb!]
     \centering
         \includegraphics[width=0.8\textwidth]{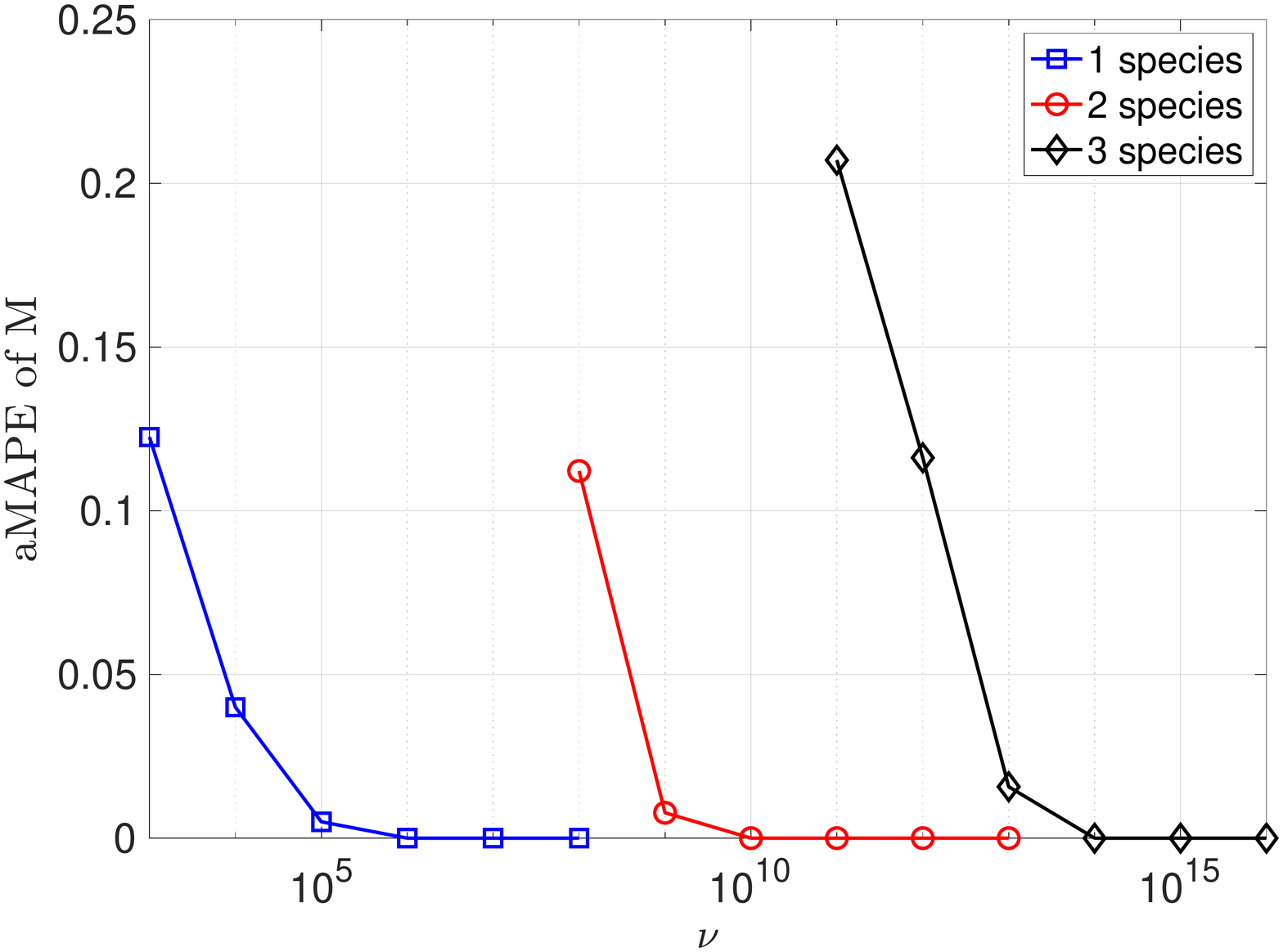}
         \caption{Given different value of $\nu$, the aMAPEs of $[\hat{M}_1,\cdots,\hat{M}_m]$ for $m=1,2,3$, $[M_1,p_1] =[8,0.1]$, $[M_2,p_2] =[10,0.2]$ and $[M_3,p_3] =[12,0.3]$. }
        \label{fig:amape}
\end{figure}

Comparing with Fig.~\ref{fig:1s}, Fig.~\ref{fig:2s} and Fig.~\ref{fig:3s}, it is noticeable that the required $\nu$ for $p_1$ to attain CRLB increases dramatically with the increasing of $m$, i.e. $\nu\approx 10^6$ for $m=1$, $\nu\approx 10^{10}$ for $m=2$ and $\nu\approx 10^{14}$ for $m=3$. Furthermore, the aMAPEs of $[\hat{M}_1,\cdots,\hat{M}_m]$ (see Eq.\ref{amape}) are plotted in Fig.~\ref{fig:amape}, where $[M_1,p_1] =[8,0.1]$, $[M_2,p_2] =[10,0.2]$ and $[M_3,p_3] =[12,0.3]$. One can see, from Fig.~\ref{fig:1s}, Fig.~\ref{fig:2s}, ,Fig.~\ref{fig:3s} and Fig.~\ref{fig:amape}, that the performance of estimating $p_j$ and $M_j$ are correlated. 

\section{Conclusion}
In this paper, we have formulated, mathematically,  the estimation of the parameters of an arbitrary number
of fluorescent species. Specifically, the convolution binomial model is presented for the underlying problem and then the exact MLE is derived. In order to resolve the intractability of the MLE manifest in the convolutional property and the high-dimensionality of the  parameters, a version of the  EM algorithm incorporating the method of moment for choosing the initial guess is proposed. The simulation results with different number of species have demonstrated the efficiency of the algorithm by comparing with the derived CRLB. 

We also found that the values of the parameters, the number of emitters and their probabilities, have impact on the performance of the estimation: the closer the probabilities are, the worse the performance is, and a larger summed numbers of emitters degrades  the performance. 

Our work provides a preliminary study and demonstrates the possibility for the estimation of an arbitrary number of fluorescent species and gives insights into the relationship  between performance of the estimator and the values of the parameters, which improves  understanding of the problem. In  future work, we will seek  (1)  to reduce  the required number of experiments while maintaining estimator performance and  (2) to improve  computational efficiency.  

\section*{Acknolwedgements}

The authors would like to thank Brett C. Johnson for his great help in the visualization of the study. This work is funded by the Air Force Office of Scientific Research (FA9550-20-1-0276). ADG also acknowledges funding from the Australian Research Council (CE140100003 and FT160100357). VVY acknowledges partial support from the National Science Foundation (NSF) (CMMI-1826078), the Air Force Office of Scientific Research (AFOSR) (FA9550-20-1-0366, FA9550-20-1-0367), DOD Army Medical Research (W81XWH2010777), the National Institutes of Health (NIH) (1R01GM127696, 1R21GM142107, 1R21CA269099), the Cancer Prevention and Research Institute of Texas (CPRIT) (RP180588). This material is also based upon work supported by the NASA, BARDA, NIH, and USFDA, under Contract/Agreement No. 80ARC023CA002E.

\bibliography{references}  

\end{document}